\documentclass[pra,twocolumn,groupedaddress,showpacs]{revtex4}

\usepackage{amsmath,amssymb}
\usepackage{epsfig}

\renewcommand{\phi}{\varphi}

\begin{document}
\bibliographystyle{apsrev}
%


\title{Critical Velocity of Superfluid Flow past Large Obstacles in Bose-Condensates}

\author{J. S. Stie\ss berger}
\author{W. Zwerger}
\affiliation{Sektion Physik, Universit\"at M\"unchen, Theresienstra\ss e 37, 80333 M\"unchen, Germany}


\date{\today}
\begin{abstract}
By considering the stability of potential flow of a superfluid around large obstacles of size $R$, we derive an analytical result for the critical velocity which is of order $v_c \sim \hbar / mR$, scaling inversely with obstacle size, in contrast to what is obtained from a Landau criterion. Our results are compared with numerical solutions of the Gross-Pitaevskii equation and with recent measurements of the critical velocity in Bose-Einstein condensates of dilute atomic gases.
\end{abstract}
%
%
\maketitle
%


Dissipationless flow for small, finite velocities is a basic and, originally, defining property of superfluids \cite{noz2}. For Bose-Einstein-condensates in dilute atomic gases, this phenomenon was recently investigated by stirring the condensate with a blue detuned laser beam \cite{ram,ono}. The associated time dependent repulsive potential of macroscopic size $R$, much larger than the healing length $\xi$, gives rise to a finite dissipation only above a critical velocity $v_c$, where superfluidity breaks down. The rate of dissipation can be obtained either by measuring the amount of heating via the resulting depletion of the condensate \cite{ram} or, more directly,  by observing the asymmetry in density associated with a finite pressure difference across the moving object \cite{ono}. The latter phenomenon is in fact completely analogous to the current induced asymmetry around defects in a Fermi liquid, as predicted long ago by Landauer as the basic microscopic origin of residual resistance \cite{land,zwe}. The measured values of $v_c$ are found to be a small fraction $v_c \approx (0.1 - 0.25) c_s$ of the sound velocity at the trap center with a numerical factor, which appears to increase with density \cite{ono}. These results are in rough agreement with numerical solutions of the Gross-Pitaevskii equation (GPE) for a moving cylindrical perturbation in a homogeneous \cite{fri,win} or inhomogeneous \cite{jac2} condensate. In the homogeneous case, for a fixed cylinder size, the critical  velocity is around $0.45 c_s$. Qualitatively this result may be understood in the spirit of the Landau criterion \cite{noz2}, by arguing that there is no dissipation as long as the local velocity $\boldsymbol{u}(\boldsymbol{x})$ at any point remains below $c_s$. For an incompressible flow, $\boldsymbol{u}(\boldsymbol{x})$ has its maximal value, which is twice the object speed, at the side of the cylinder $|\boldsymbol{x}| = R$, $\theta = \pm \pi/2$ \cite{lan6}. Since the local sound velocity is reduced by the depletion of the actual condensate density close to the obstacle, a value $v_c$ smaller than $c_s/2$ (or $2/3\  c_s$ for a sphere) but {\em independent} of $R$ is expected from this type of argument. In an inhomogeneous situation like that for atomic condensates in a harmonic trap, the critical velocity will be lowered further because the density decreases away from the trap center.

Our aim in the present work is to derive an analytical criterion for the critical velocity of superfluid flow around macroscopic obstacles of size $R \gg \xi$ by considering the linear stability of the dissipationless potential flow below $v_c$ towards the generation of vortices. For a strongly repulsive potential, $v_c$ is of order $ 5 \hbar / m R$ i.e. independent of the density and inversely proportional to the obstacle size. Since $\hbar / m \xi = \sqrt{2} c_s$, the critical velocity coincides with a fraction of the sound velocity for obstacle sizes $R \approx (10 - 20) \xi$, which are in fact typical values considered both experimentally and in the numerical simulations. As pointed out by Nozi\`eres and Pines \cite{noz2}, the prediction $v_c \sim \hbar / mR$ is a generic result for the critical velocity due to vortex generation, as shown e.g. by Feynman's estimate $v_c \approx \frac{\hbar}{mD} \ln \frac{D}{\xi} \ll c_s$ \cite{fey} for superfluid flow in a long channel with diameter $D \gg \xi$ \cite{fet}. A very similar result for a cylindrical trap was obtained very recently by Crescimanno et al \cite{cre}.

The dynamics of a dilute, weakly interacting Bose-Einstein-condensate (BEC) is accurately described by the GPE for the complex condensate wave function $\Phi$ \cite{dal}. Since we are considering a bulk situation, it is convenient to normalize $\Phi$ according to $\int |\Phi|^2 d^3 x = N$. In the rest frame of the obstacle, where the problem is stationary, the energy functional ${\cal F}[\Phi]$ of a homogeneous condensate moving with velocity $\boldsymbol{v}$, has the standard form
\begin{equation}\label{eq:energy}
{\cal F}[\Phi] ={\cal E}[\Phi]  - \boldsymbol{v} \cdot \boldsymbol{\cal P}[\Phi] \ .
\end{equation}
Here
\begin{equation}
  \label{eq:E}
{\cal E}[\Phi] = \int d^{d} x \left[ \frac{\hbar^2}{2m} |\boldsymbol{\nabla} \Phi|^2 + \frac{g}{2} |\Phi|^4 \right]
\end{equation}
is the condensate energy functional in its rest frame and
\begin{equation}
  \label{eq:P}
\boldsymbol{\cal P}[\Phi] = \int d^d x \frac{i \hbar}{2} \left[ \Phi \boldsymbol{\nabla} \Phi^{*} - \Phi^{*} \boldsymbol{\nabla} \Phi \right]
\end{equation}
is the momentum functional of the moving condensate. For dilute gases, the interaction parameter $g = 4\pi \hbar^2 a / m$ is determined by the scattering length $a$, which is assumed to be positive. Splitting the condensate wave function $\Phi(\boldsymbol{x}) = f(\boldsymbol{x}) e^{i \phi(\boldsymbol{x})}$ into magnitude and phase, the energy functional (\ref{eq:energy}) takes the form
\begin{eqnarray}
  \label{eq:F(fphi)}
{\cal F}[f, \phi] & = & \int d^d x \left[ \frac{\hbar^2}{2m} \left( (\boldsymbol{\nabla} f)^2 + f^2 (\boldsymbol{\nabla} \phi)^2 \right)  \right. \nonumber \\ 
&+ & \left. \frac{g}{2} f^4 - \hbar \boldsymbol{v}\cdot (\boldsymbol{\nabla} \phi) f^2 \right] \ .
\end{eqnarray}
In order to calculate the critical velocity, we determine the limit beyond which small fluctuations around the dissipationless potential flow below $v_c$ are no longer stable. Since $\mbox{rot} \boldsymbol{u}$ vanishes identically below $v_c$, the condensate velocity relative to the obstacle at rest can quite generally be written in the form
\begin{equation}
  \label{eq:velocity}
  \boldsymbol{u}(\boldsymbol{x}) = \frac{\hbar}{m} \boldsymbol{\nabla} \phi(\boldsymbol{x}) + \boldsymbol{v} \ ,
\end{equation}
where the term $\hbar / m \boldsymbol{\nabla} \phi$ accounts for the backflow induced by the presence of the obstacle. Far from the obstacle, where the density $f^2$ is constant, the backflow is necessarily of a dipolar form
\begin{equation}
  \label{eq:vel_of_obstacle}
\frac{\hbar}{m} \boldsymbol{\nabla} \phi = - A_d \frac{\left( d \hat{\boldsymbol{n}} (\boldsymbol{v} \cdot \hat{\boldsymbol{n}}) - \boldsymbol{v} \right)}{r^d} \ ,
\end{equation}
since the dipole is the only function obeying both conditions of vanishing vorticity $\mbox{rot} \boldsymbol{u} = 0$ and stationarity $0 = \mbox{div} f^2 \boldsymbol{u} \approx f^2 \mbox{div} \boldsymbol{u}$.
Here $\hat{\boldsymbol{n}}$ is the unit vector in the direction of $\boldsymbol{x}$ and $d =2,3$ the relevant dimensionality ($d=2$ for a moving cylinder, $d=3$ for a sphere). 
For a weak obstacle potential $V_B \ll \mu$, the response of the superfluid to the moving object can be treated in first order perturbation theory \cite{noz1}. The resulting dipole strength $A_d \sim \frac{1}{n} \frac{\partial n}{\partial \mu} \cdot V(\boldsymbol{q} = 0)$ is determined by the bulk compressibility $n \kappa = \frac{1}{n} \frac{\partial n}{\partial \mu} = (m c_{s}^{2})^{-1}$ and the Fourier transform of the potential at vanishing wave vector. For a potential of size $R$ this leads to a behaviour $A_d \sim V_B / \mu \cdot R^d$. In the experiments, however, the repulsive potential is rather strong $V_B \gg \mu$ and, in particular, we want to investigate the hard core limit $V_B \rightarrow \infty$, where perturbation theory definitely fails. In this regime the obstacle is equivalent to the boundary condition $\boldsymbol{u}\cdot \hat{\boldsymbol{n}} |_R =0$ of vanishing normal velocity at $| \boldsymbol{x}| = R$. Provided the backflow has the form given in (\ref{eq:vel_of_obstacle}) for arbitrary values of $r$, this immediately fixes  $A_d = R^d / (d-1)$. With this approximation, (\ref{eq:vel_of_obstacle}) is just the flow pattern of an incompressible, ideal classical fluid around an obstacle, which has a vanishing drag force in a stationary situation $\boldsymbol{v} = \mbox{const.}$ \cite{lan6}. To account for the depletion in density due to the finite compressibility of a real BEC, we use the following Ansatz
\begin{equation}
  \label{eq:tanh_form}
f(r) = \begin{cases}
\sqrt{\frac{\mu}{g}} \tanh\left(\frac{r- \delta}{\sqrt{2} \xi} \right) & \quad \mbox{for}\ r > R \\
\sqrt{\frac{\mu}{g}} {\cal A} \exp \left( \kappa_{\mu} (r- R) \right) & \quad \mbox{for}\ r < R\ ,
\end{cases}
\end{equation}
with $\xi = \hbar / \sqrt{2m \mu}$, $\kappa_{\mu} = \sqrt{2m(V_B - \mu)/ \hbar^2}$ and obstacle potential $V_B$. Equation (\ref{eq:tanh_form}) combines the standard one dimensional solution of the GPE at an infinite potential step with a one-particle wave function which decays exponentially below the barrier $V_B$. The latter is a solution of the GPE without the $\Phi^3$ term and thus is valid in the limit $V_B \gg \mu$, where the external potential due to the obstacle is much larger than the mean field interaction energy. Matching the logarithmic derivative of $f(r)$ at $r=R$, leads to the following expressions for $\delta$ and ${\cal A}$
\begin{eqnarray}
  \label{eq:match_of_f}
  {\cal A} & = & \frac{\xi \kappa_{\mu}}{\sqrt{2}} \left(\sqrt{1+ \frac{2}{\xi^2 \kappa^2_{\mu}}} -1 \right)\ , \\
\delta & = & - \frac{\xi}{2} \ln \left(\frac{1 + {\cal A}}{1 - {\cal A}} \right) + R\ .
\end{eqnarray}
Equation (\ref{eq:tanh_form}) smoothly interpolates on the scale $\xi$ between the bulk condensate density $n = \mu / g$ far from the obstacle to essentially zero at its center $r=0$ (note $R \gg \xi$). The Ansatz (\ref{eq:tanh_form}) thus describes a ``soft'', penetrable obstacle as realized experimentally. Although (\ref{eq:tanh_form}) is not an exact solution of the 3-dimensional GPE, comparison to numerical solutions shows that the approximation is excellent \cite{nor}.

\begin{figure}[htbh]
  \begin{center}
    \epsfig{file=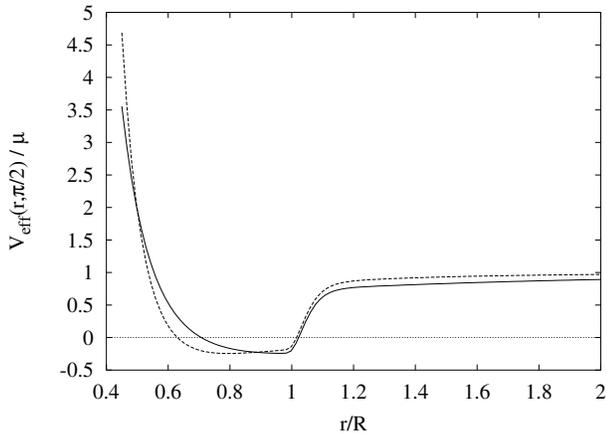, width=8.5cm}
    \caption{Effective potential $V_{eff}(r,\theta = \pi/2, v= 0.7 c_s)$ for $V_B = 10 \mu$ and $R = 20 \xi$. The solid line is the potential in $d=3$ and the dashed line in $d=2$.}
    \label{fig:potential}
  \end{center}
\end{figure}

To calculate the critical velocity of the superfluid flow, we follow a method similar to that used to determine the stability of electrical supercurrents in a wire by Langer and Ambegaokar \cite{lan}. We consider small deviations around the approximate stationary point $\bar{\Phi} = f e^{i\phi}$ described by equations (\ref{eq:vel_of_obstacle}) and (\ref{eq:tanh_form}) which are conveniently written in the form
\begin{equation}
\Phi = (f(\boldsymbol{r}) + w(\boldsymbol{r})) \exp(i \phi(\boldsymbol{r})) \ .
\end{equation}
For a general complex function $w(\boldsymbol{r})$, this Ansatz includes both amplitude and phase fluctuations around the stationary solution with phase $\phi(\boldsymbol{r})$. Since the potential flow around the obstacle described by $\bar{\Phi}$ is a stationary point of the energy functional (\ref{eq:energy}), the expansion of ${\cal F}[\Phi]$ up to second order in $w$
\begin{equation}
  \label{eq:energy_expansion}
  {\cal F}[\Phi] = {\cal F}[\bar{\Phi}] + {\cal Q}_{\bar{\Phi}}[w] + \dots \ ,
\end{equation}
yields a quadratic form ${\cal Q}_{\bar{\Phi}}[w]$. The eigenvalues of ${\cal Q}_{\bar{\Phi}}[w]$ give the characteristic curvature of ${\cal F}[\Phi]$ at $\bar{\Phi} = f e^{i\phi}$ in function space. In order for ${\cal F}$ to have a local minimum at $\bar{\Phi}$, which means that the solution is stable, all the eigenvalues of ${\cal Q}_{\bar{\Phi}}[w]$ must be positive. The lowest velocity $v$ for which ${\cal Q}$ has an eigenfunction with negative eigenvalue thus gives the critical velocity $v_c$ for the onset of dissipation in the BEC \footnote{In the case of a uniform condensate, this approach leads to an eigenvalue spectrum which is positive unless $v > c_s$.}. The eigenvalue equation $\frac{\delta {\cal Q}}{\delta w^{*}} = \lambda w$ can be written in the form
\begin{eqnarray}
  \label{eq:eigenvalue}
 & -& \frac{\hbar^2}{2m} \nabla^2 w + \left(\frac{\hbar^2 (\boldsymbol{\nabla}\phi)^2}{2m} - \hbar  \boldsymbol{\nabla}\phi \cdot \boldsymbol{v} \right)w \nonumber \\
&+& 2 f^2 gw + f^2g w^{*} = \lambda w \ .
\end{eqnarray}
Similar to the situation discussed in \cite{lan}, it turns out that a purely complex $w$ yields the smallest eigenvalue and thus the smallest critical velocity. Physically this means that the instability is driven by phase fluctuations associated with vortex generation, as also found numerically \cite{jac1}. Using $w = i \tilde{w}$, the eigenvalue equation (\ref{eq:eigenvalue}) can be written as a Schr\"odinger equation
\begin{equation}
  \label{eq:eigen}
 - \frac{\hbar^2}{2m} \nabla^2 \tilde{w} + V_{eff}(r,\theta,v) \tilde{w} = \lambda \tilde{w} \ ,
\end{equation} 
with an effective potential which is determined by inserting the approximations (\ref{eq:vel_of_obstacle}) and (\ref{eq:tanh_form}) in equation (\ref{eq:eigenvalue}). The effective potential thus has the following form in $d$ dimensions
\begin{eqnarray}
\label{eq:eff_potential}
V_{eff}(r,\theta,v) & = & \frac{m v^2}{2} \left[ \left(\frac{A_d}{r^d} \right)^{2} \left\{ (d^2-2d)\cos^2 \theta +1 \right\} \right. \nonumber \\
& + & \left. \frac{2 A_d}{r^d} \left\{d \cos^2 \theta -1 \right\} \right] + g f^2(r) \ .
\end{eqnarray}
Examining the effective potential, we find that an attractive well is formed around $\theta = \pm \pi /2$, where bound states with negative eigenvalue may develop (see Fig. \ref{fig:potential}). The depth of this potential well increases with the velocity $v$, and thus there is a critical velocity $v_c$ beyond which equation (\ref{eq:eigen}) has a negative eigenvalue.
In agreement with numerical simulations \cite{jac1}, $\theta = \pm \pi /2$ is the position at the side of the obstacle where vortices are generated at the breakdown of superfluidity, i.e. it is the barrier for vortex creation which vanishes at $v_c$.

In principle, the velocity at which the lowest eigenvalue of the Schr\"odinger equation with the anisotropic potential (\ref{eq:eff_potential}) becomes negative may be found numerically. In order to obtain an analytical result which displays the relevant parameter dependence of the critical velocity, however, it is more convenient to use  a simple semiclassical estimate for the number of eigenstates in the effective potential \cite{sim}
\begin{equation}
  \label{eq:semiclassicalN}
  N_{sc}(v) = \left(\frac{\sqrt{2m}}{2 \pi \hbar}\right)^{d} V_d \int_{V_{eff} < 0} d^{d} x \left| V_{eff}(r,\theta,v) \right|^{d/2} \ ,
\end{equation}
where $V_d$ is the volume of the unit sphere in $d$ dimensions. Equation (\ref{eq:semiclassicalN}) gives the number of states which have negative eigenvalue and is just the classical phase space volume of the region where $V_{eff} < 0$ in units of $(2 \pi \hbar)^{d}$. The condition $N_{sc}(v_c) = 1$ thus gives a simple semiclassical approximation for the critical velocity.

Introducing reduced units $\tilde{R} = R/ \xi$ and $V_B/ \mu$, the condition $N_{sc} =1$ takes the following form for the cylindrical obstacle
\begin{eqnarray}
  \label{eq:RESULT_cyl}
  1 & = & \frac{\tilde{R}^2}{8 \pi} \int_{0}^{2 \pi} d \theta \int_{0}^{\infty} \rho  d \rho \left| \min \left[0, \left(\frac{v_c}{c_s} \right)^2 \left\{ \frac{1}{\rho^4}  \right. \right. \right. \nonumber \\
& + &\left. \left. \left.  \frac{2}{\rho^2} \left( 2 \cos^2 \theta - 1 \right)  \right\} +  2 \tilde{f}^2(\rho) \right] \right| \ .
\end{eqnarray}
For the spherical perturbation the corresponding condition for $v_c$ reads
\begin{eqnarray}
  \label{eq:RESULT_sphere}
& &  1 = \frac{\tilde{R}^3}{6 \sqrt{2} \pi} \int_{0}^{\pi} \sin \theta  d \theta \int_{0}^{\infty} \rho^2  d \rho   \left| \min \left[0, \left(\frac{v_c}{c_s} \right)^2 \left\{ \frac{1}{4 \rho^6} \right. \right. \right. \nonumber \\
& &\times   \left. \left. \left. \left(3 \cos^2 \theta + 1\right) + \frac{1}{\rho^3} \left(3 \cos^2 \theta -1\right) \right\} + 2 \tilde{f}^2(\rho) \right] \right|^{3/2}
\end{eqnarray}
In (\ref{eq:RESULT_cyl}, \ref{eq:RESULT_sphere}) $\tilde{f}$ is the function (\ref{eq:tanh_form}) in reduced units
\begin{equation}
  \label{eq:tanh_form_red}
\tilde{f}(\rho) = \begin{cases}
\tanh\left(\frac{\tilde{R}}{\sqrt{2}} \left(\rho - \frac{\delta / \xi}{\tilde{R}} \right) \right) & \mbox{for}\ \rho > 1  \\
{\cal A} \exp \left(\tilde{R} \sqrt{\frac{V_B}{\mu} - 1} (\rho - 1) \right) & \mbox{for}\ \rho < 1\ .
\end{cases}
\end{equation}

\begin{figure}[htbp]
  \begin{center}
    \epsfig{file=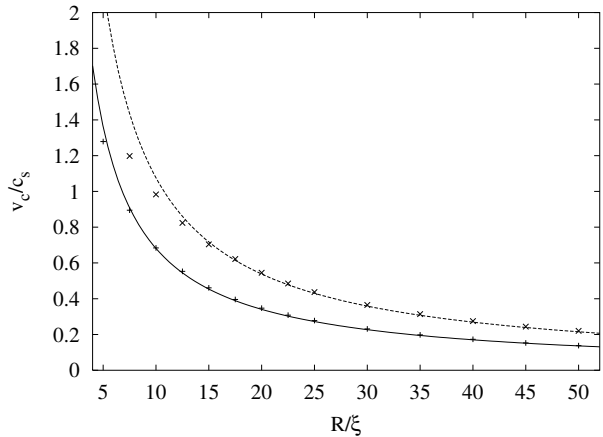, width=8.5cm}
    \caption{Critical velocity $v_c$ as a function of the obstacle radius for a potential $V_B = 10 \mu$. The upper curve presents the moving cylinder and the lower one the moving sphere. The results obtained from (\ref{eq:RESULT_cyl}) and  (\ref{eq:RESULT_sphere}) are shown as crosses, the lines exhibit the $1/R$ dependence.}
    \label{fig:crit_vel}
  \end{center}
\end{figure}

Equations (\ref{eq:RESULT_cyl}) and (\ref{eq:RESULT_sphere}) are the main results of this work, providing a simple criterion for the critical velocity of a moving cylinder and sphere.
Evidently the ratio $v_c / c_s$ depends only on the two dimensionless parameters $V_B / \mu$ and $R / \xi$. In the limit $R / \xi \gg 1$ and $V_B / \mu\gg 1$, the function (\ref{eq:tanh_form_red}) converges to the step function $\theta(\rho - 1)$. In this limit the integrals in (\ref{eq:RESULT_cyl}) and (\ref{eq:RESULT_sphere}) are independent of $R / \xi$ and $V_B / \mu$. The dependence of the critical velocity on the obstacle radius is therefore simply
\begin{equation}
  \label{eq:uc_dependence}
  v_c \sim \frac{\hbar}{mR} \ ,
\end{equation}
The proportionality factor for the cylinder and sphere can be easily evaluated giving $v_c =7.61 \hbar / mR$ and $v_c= 4.82 \hbar / mR$ respectively. The precise numerical factor is in agreement with the lines shown in Fig. \ref{fig:crit_vel}. 
It will change in an exact solution of the Schr\"odinger equation (\ref{eq:eigen}), without however, affecting the basic result $v_c  = \mbox{const. } \hbar / mR$. As shown in Fig. \ref{fig:crit_vel} the numerical integration of (\ref{eq:RESULT_cyl}, \ref{eq:RESULT_sphere}) gives critical velocities which very closely obey the $1/R$-scaling (\ref{eq:uc_dependence}) for radii larger than about $10 \xi$ or $15 \xi$ for the sphere or the cylinder. For small radii $R \lesssim 7 \xi$, the critical velocity becomes larger than the sound velocity. In this regime our approximations are no longer valid and a more microscopic theory is necessary to determine $v_c$. Choosing different values for $V_B / \mu$, it turns out that the result shown in Fig. \ref{fig:crit_vel} remains unchanged down to $V_B \approx 5 \mu$, beyond which $v_c$ starts to increase, in agreement with the behaviour found numerically \cite{jac2}. Numerical work on a moving cylinder in a homogeneous BEC \cite{fri,win} observed a critical velocity of about $0.45 c_s$ for $R > \xi$. No dependence of the results on the obstacle radius was reported, however. Simulations for a trapped BEC and an obstacle with Gaussian object potential reported a critical velocity between $0.55 c_s$ and $0.3 c_s$ \cite{jac2}, close to the typical values for $v_c$ found in this work.

In the experiments \cite{ram,ono} the blue detuned laser beam which served as a cylindrical obstacle had a radius of $R \approx 20 \xi$ and a potential barrier $V_B = (6.4 - 8) \mu$. The measured critical velocity in the trapped BEC was about $(0.1 - 0.25) c_s$. 
For the values of ref. \cite{ram} our approach yields a critical velocity $v_c = 0.51 c_s$ for a perturbation moving uniformly in a homogeneous condensate. Compared with the experiments, this is about a factor of two larger. Now, as has been found in recent numerical simulations of the GPE \cite{jac2}, the measured critical velocity is lower than in the bulk for two reasons: First of all, in the trapped BEC, the obstacle touches regions of the condensate with lower density and thus a lower velocity of sound. Accordingly vortices will appear first in the outer regions of the condensate and penetrate towards the trap center. Secondly, due to the oscillatory movement, the obstacle creates its own wake, thus lowering the critical velocity further. 
In the experiment \cite{ono} the critical velocity was measured for different condensate densities $n$ at fixed $R$. In contrast to our result (\ref{eq:uc_dependence}), the critical velocity seems to scale with $c_s \sim \sqrt{n}$ or even more strongly with density. It is possible that this discrepancy is caused by our incompressible fluid approximation for the strength $A_d$ of the dipolar backflow. Clearly a more microscopic calculation of $A_d$ for strong scattering is required to clarify the situation and also experiments, in which the effective obstacle size $R$ is varied at fixed density.

In conclusion, we have derived a simple analytical result for the critical velocity of a macroscopic moving obstacle in a BEC. In particular it has been shown that the critical velocity scales inversely with the obstacle size, a prediction which should easily be checked experimentally.

We thank W. Ketterle for helpful comments.
%
%
%

\end{document}